\documentclass[12pt]{amsart}
\usepackage[T1]{fontenc}
\usepackage[english]{babel}
\usepackage{color}
\usepackage{mathtools}
\usepackage{graphicx}
\usepackage{amssymb}
\usepackage{mathabx}
\usepackage{tikz}
\usepackage{tikz-cd}
\usepackage{hyperref}
\usepackage{dsfont,soul}
\usepackage{enumitem}
\graphicspath{{graphics/}}
\usepackage{geometry}
\setlist[enumerate]{labelsep=*, leftmargin=1.5pc,
topsep=1ex plus0.5ex minus0.2ex,
itemsep=1ex plus0.5ex minus0.2ex,
font=\rmfamily,
font=\upshape}
\setlist[itemize]{labelsep=*, leftmargin=1.5pc,
topsep=1ex plus0.5ex minus0.2ex,
itemsep=1ex plus0.5ex minus0.2ex,
font=\rmfamily,
font=\upshape}
\newtheorem{thm}{Theorem}[section]

\newtheorem{lem}[thm]{Lemma}
\newtheorem{pro}[thm]{Proposition}

\theoremstyle{definition}
\newtheorem{defn}[thm]{Definition}

\newtheorem{rem}[thm]{Remark}
\numberwithin{equation}{section}
\renewcommand{\Re}{\operatorname{Re}}
\renewcommand{\Im}{\operatorname{Im}}
\newcommand{\bC}{\mathbb C}

\newcommand{\bN}{\mathbb N}
\newcommand{\bR}{\mathbb R}
\newcommand{\cA}{\mathcal A}
\newcommand{\cD}{\mathcal D}
\newcommand{\cE}{\mathcal E}

\newcommand{\cK}{\mathcal K}
\newcommand{\cL}{\mathcal L}
\newcommand{\cN}{\mathcal N}
\newcommand{\cP}{\mathcal P}
\newcommand{\cQ}{\mathcal Q}

\newcommand{\id}{\mathds{1}} % identity matrix

\DeclareMathOperator{\ii}{i} % imaginary unit
\DeclareMathOperator*{\argmin}{argmin}

\DeclareMathOperator{\tr}{tr}
\DeclareMathOperator{\cl}{cl}
\DeclareMathOperator{\supp}{supp}
\begin{document}
\selectlanguage{english}
\title{A variation principle for ground spaces}
\author{Stephan Weis}
\begin{abstract}
The ground spaces of a vector space of hermitian matrices, partially 
ordered by inclusion,  form a lattice constructible from top to 
bottom in terms of intersections of maximal ground spaces. In this 
paper we characterize the lattice elements and the maximal lattice 
elements within the set of all subspaces using constraints on 
operator cones. Our results contribute to the geometry of quantum 
marginals, as their lattices of exposed faces are isomorphic to the 
lattices of ground spaces of local Hamiltonians. 
\end{abstract}
%
% in use 
%
% Primary:
% 52A20 Convex sets in $n$ dimensions
% 52B05 Combinatorial properties (number of faces, shortest paths, etc.)
% 51D25 Lattices of subspaces
% 47L07 Convex sets and cones of operators
% 47A12 Numerical range, numerical radius
% 81P16 Quantum state spaces, operational and probabilistic concepts
% Secondary:
% 62F30 Inference under constraints
% 94A17 Measures of information, entropy
%
% not in use
%
% 62H20 Measures of association (correlation, canonical correlation, etc.)
% 81P45 Quantum information, communication, networks
%
\maketitle
\thispagestyle{empty}
\pagestyle{myheadings}
\markleft{\hfill A variation principle for ground spaces\hfill}
\markright{\hfill S.\ Weis \hfill}
%
%%%%%%%%%%%%%%%%%%%%%%%%%%%%%%%%%%%%%%%%%%%%%%%%%%%%%%%%%%%%%%%%%%%%%%%%%%%%
%%%%%%%%%%%%%%%%%%%%%%%%%%%%%%%%%%%%%%%%%%%%%%%%%%%%%%%%%%%%%%%%%%%%%%%%%%%%
%%%%%%%%%%%%%%%%%%%%%%%%%%%%%%%%%%%%%%%%%%%%%%%%%%%%%%%%%%%%%%%%%%%%%%%%%%%%
%%%%%%%%%%%%%%%%%%%%%%%%%%%%%%%%%%%%%%%%%%%%%%%%%%%%%%%%%%%%%%%%%%%%%%%%%%%%
%%%%%%%%%%%%%%%%%%%%%%%%%%%%%%%%%%%%%%%%%%%%%%%%%%%%%%%%%%%%%%%%%%%%%%%%%%%%
%
%
\section{Introduction}
\label{sec:intro}
\par
The variation principle \cite{Griffiths2005,SakuraiNapolitano2011} for 
estimating the ground energy $E_0$ (smallest eigenvalue) of a Hamiltonian 
$h$ (hermitian matrix) asserts that 
\begin{equation}\label{eq:1}\textstyle
E_0 \leq \langle\psi|h|\psi\rangle
\end{equation}
for all pure states $|\psi\rangle$. This means that $E_0$ is the smallest 
expected energy a quantum system can have in any pure state $|\psi\rangle$ 
if $h$ is the energy operator. The upper bound on $E_0$ is 
mathematically trivial but computationally very useful. Nevertheless, 
the estimation of $E_0$ for the class of {\em $k$-local Hamiltonians}, 
describing the energy of a many-body system without interactions 
between more than $k$ units, is a hard problem already for $k=2$
\cite{Kitaev-etal2002,Kempe-etal2006,CubittMontanaro2016}. Geometric 
approaches to the {\em local Hamiltonian problem} of estimating $E_0$
are discussed since the 1960's. A basic idea is that $E_0$ is the 
displacement of the supporting hyperplane with normal vector $h$ 
touching the convex set $\cD_{(k)}$ of {\em $k$-body marginals}. 
\par
Similar to the local Hamiltonian problem, the {\em quantum marginal problem} 
of deciding whether a tuple of states lies in $\cD_{(k)}$ is neither 
believed to have efficiently computable solutions \cite{Liu2006,Zeng-etal2015}. 
Many of the results pertaining to $\cD_{(k)}$ were originally obtained in the 
fermionic case (not treated here). Notably, spectral properties of marginal 
density matrices
\cite{Coleman1963,AltunbulakKlyachko2008,Sawicki-etal2013,Christandl-etal2014}
were discovered. Spectral conditions are insufficient to characterize 
$k$-body marginals of overlapping subsystems ($k>1$), for that there are 
results concerning extreme points 
\cite{Coleman1963,Erdahl1972,Ocko-etal2011,Chen-etal2012b} and 
order-theoretic results regarding ground spaces of marginal states 
\cite{Chen-etal2012a}. 
\par
In this article we study order-theoretic aspects of the convex geometry of a 
set of quantum marginals---we begin with linear images of general convex sets,
then linear images of state spaces of matrix algebras, and finally the very
special case of quantum marginals $\cD_{(k)}$. More precisely, a quantum 
mechanical system \cite{BratteliRobinson1987,AlfsenShultz2001} 
is described by a complex *-subalgebra of $M_n$ including the $n$-by-$n$ 
identity matrix $\id$. A density matrix, or {\em state}, of $\cA$ is a 
matrix of trace one in the cone $\cA^+:=\{a\in\cA:a\succeq 0\}$ of positive 
semi-definite matrices in $\cA$. The states of $\cA$ form a convex set 
$C_\cA$, called {\em state space}. 
\par
According to the variation principle (\ref{eq:1}) and using a simple 
convexity argument, the ground energy of a Hamiltonian $u\in\cA$ is
\[\textstyle
\min\{\langle \rho,u\rangle : \rho\in C_\cA \},
\]
where $\langle a,b\rangle=\tr(a^*b)$ is the Hilbert-Schmidt inner product 
of $a,b\in\cA$. Assuming $u$ lies in a linear subspace $U\subset\cA$ of 
hermitian matrices, Euclidean geometry shows 
\begin{equation}\label{eq:ground-st-energy-loc}\textstyle
\min\{\langle b,u\rangle : b\in\pi(C_\cA)\},
\qquad
u\in U,
\end{equation}
where $\pi$ is the orthogonal projection onto $U$. 
Geometrically, equation (\ref{eq:ground-st-energy-loc}) means that
the ground energy restricted to $U$ is the distance of the 
origin from the supporting hyperplane of $\pi(C_\cA)$ with inner normal 
vector $u$.
\par
A well-known aspect of $\pi(C_\cA)$ is the isomorphism between its 
exposed faces and the ground projections of $U$. An {\em exposed face} 
of $\pi(C_\cA)$ is the intersection of $\pi(C_\cA)$ with a supporting 
hyperplane, that is the subset of $\pi(C_\cA)$ at which the minimum 
(\ref{eq:ground-st-energy-loc}) is achieved for some $u\in U$. The 
{\em ground projection} $p_0(u)$ of $u$ is the spectral projection of 
$u$ corresponding to the smallest eigenvalue. The {\em ground space} of
$u$ is the image of $p_0(u)$. The set of exposed faces of $\pi(C_\cA)$, 
partially ordered by inclusion, is lattice-isomorphic to the set 
$\cP(U)$ of ground projections of $U$, partially ordered by the 
L\"owner partial ordering \cite{Weis2018}. 
\par
A discussion of normal cones of $\pi(C_\cA)$ leads to the 
Definition~\ref{def:K} of the cone
\[
K(p):=p'\cA^+p'\cap U
\]
for {\em projections} $p\in\cA$, that is $p=p^*=p^2$. Thereby, $p'=\id-p$
is the complementary projection of $p$, and $p'\cA^+p'=\{p'ap':a\in\cA^+\}$.
Notice that 
\[
K(p)=\{u\in U \mid u\succeq 0, p\preceq\ker(u)\},
\]
where $\ker(u)$ is the kernel projection of $u$. It is easy to show that if 
$\id\in U$, then the ground projection $p_0(u)$ of any $u\in U$ is greater
than or equal to any projection $p\in\cA$ satisfying $K(p)\supset K(p_0(u))$. 
Theorem~\ref{thm:char-PU-K} completes this statement to what we call 
{\em variation principle}: For every projection $p\in\cA$, 
the set of projections $q\in\cA$ satisfying $K(q)=K(p)$ has a greatest element 
$q_{\max}(p)$ and $q_{\max}(p)\in\cP(U)$.
\par
The variation principle shows that the decision problem of whether a projection 
$p\in\cA$ lies in $\cP(U)$ is equivalent to that of whether $p=q_{\max}(p)$. 
One way to compute $q_{\max}(p)$ is to pick a relative interior point $u$ of 
$K(p)$ and compute its ground projection $p_0(u)$, as $q_{\max}(p)=p_0(u)$ 
follows from results concerning normal cones of state spaces \cite{Weis2011}. 
This raises the question, not teated here, of whether a relative interior point 
of $K(p)$ can be efficiently computed.
\par
The computation of $\cP(U)$ can be put down to that of the maximal elements of 
$\cP(U)\setminus\{\id\}$, known as {\em coatoms}, as the latter generate 
$\cP(U)$ in terms of infima \cite{Weis2018}. Clearly, a projection $p$ of 
$\cA$ of rank $n-1$ is a coatom if and only if $p'\in U$, that is if $K(p)$ is 
a ray. Theorem~\ref{thm:ray-coatoms}(1) proves for any $p\in\cP(U)$ that $p$
is a coatom if and only if $K(p)$ is a ray. The interesting part of the 
theorem, that $p\in\cP(U)$ is not a coatom if $\dim(K(p))>1$, is exploited in 
Section~\ref{sec:Composite}. Infima of projections are studied, for example, in 
\cite{KuboAndo1980,Gheondea-etal2005,BoettcherSpitkovsky2010}.
\par
The article is organized as follows.
Section~\ref{sec:var-lift} proves a variation principle for convex sets. 
Sections~\ref{sec:lattices-projections} and~\ref{sec:lattices} recall 
lattice isomorphisms related to the state space and its linear images.
Section~\ref{sec:var-gs} translates the variation principle to ground spaces. 
Section~\ref{sec:coatoms-gs} characterizes coatoms. 
Section~\ref{sec:simple-non-commutative} discusses a
non-commutative example.
Sections~\ref{sec:Composite} and~\ref{sec:example-3bit} deal with local 
Hamiltonians. Section~\ref{sec:conclusion} is a conclusion.
\begin{rem}[Complexity of $\cD_{(k)}$]
The {\em semi-definite extension complexity} 
\cite{Fiorini-etal2012,Averkov-etal2017} of $\cD_{(k)}$ is the minimal 
dimension $d$ for which $\cD_{(k)}$ is a linear image of an affine 
section of $M_d^+$. The local Hamiltonian problem being hard for 
$k\geq 2$, the existence of efficient algorithms 
\cite{NesterovNemirovskij2001} for linear optimization on $M_d^+$ suggests
that the complexity $O(2^N)$ of $\cD_{(k)}$ for $N$ qubits is unlikely to 
be polynomial in $N$. Because of the gap from $O(2^N)$ to the polynomial 
$\dim(\cD_{(k)})=\Theta(N^k)$, the set $\cD_{(k)}$ has a much richer 
convex geometry than sections and projections of $M_d^+$ can have for $d$ 
of order $\sqrt{N^k}$.
\end{rem}
\begin{rem}[Topology of $\cP(U)$]
In addition to contributing to the geometry of $\pi(C_\cA)$, we hope that 
the results will ultimately enable us to discuss the topology of $\cP(U)$. 
If the algebra $\cA$ is non-commutative, then $\cP(U)$ may not be closed 
in the norm topology, because of discontinuities of the maximum-entropy 
inference map $\pi(C_\cA)\to C_\cA$ under linear constraints 
\cite{WeisKnauf2012}. For example, $\cP(U_{(2)})$ is not norm closed for 
three qubits \cite{Rodman-etal2016}. For a many-body
system, the discontinuity of the inference map $\cD_{(k)}\to C_\cA$ is 
equivalent \cite{Weis2014,Weis-etal2015,Rodman-etal2016} to that of a 
correlation quantity
\cite{Amari2001,AyKnauf2006,Zhou2008,Ay-etal2011,Rauh2011,Niekamp-etal2013,
Guehne-etal2017} closely related to the ``irreducible correlation'' 
first introduced in \cite{Linden-etal2002}. For $k=1$, the quantity is the
mutual information or multi-information, which is continuous and which 
measures the total correlation \cite{Groisman-etal2005}.
\end{rem}
%
%%%%%%%%%%%%%%%%%%%%%%%%%%%%%%%%%%%%%%%%%%%%%%%%%%%%%%%%%%%%%%%%%%%%%%%%%%%%
%%%%%%%%%%%%%%%%%%%%%%%%%%%%%%%%%%%%%%%%%%%%%%%%%%%%%%%%%%%%%%%%%%%%%%%%%%%%
%%%%%%%%%%%%%%%%%%%%%%%%%%%%%%%%%%%%%%%%%%%%%%%%%%%%%%%%%%%%%%%%%%%%%%%%%%%%
%%%%%%%%%%%%%%%%%%%%%%%%%%%%%%%%%%%%%%%%%%%%%%%%%%%%%%%%%%%%%%%%%%%%%%%%%%%%
%%%%%%%%%%%%%%%%%%%%%%%%%%%%%%%%%%%%%%%%%%%%%%%%%%%%%%%%%%%%%%%%%%%%%%%%%%%%
%
\section{A variation principle for pre-images of exposed faces}
\label{sec:var-lift}
\par
We show that the pre-images of exposed faces of a projection of a 
convex set $C$ are the greatest exposed faces under a normal cone 
constraint. 
\begin{defn}
A {\em closure operation} \cite{Aigner1997,Birkhoff1967} on a set 
$I$ is an operator $2^I\to 2^I$, $X\mapsto\cl(X)$ on the subsets 
of $I$ such that for all $X,Y\subset I$ we have 
$X\subset\cl(X)$ (extensive),
$\cl(\cl\,X)=\cl(X)$ (idempotent),
and $X\subset Y\implies\cl(X)\subset\cl(Y)$ (isotone).
Subsets $X\subset I$ with $X=\cl(X)$ are called 
{\em closed sets} with respect to $\cl$.
\end{defn}
\begin{lem}\label{lem:1}
Let $\cl$ be a closure operation on a set $I$. Let ${\mathcal S}\subset 2^I$ 
contain all closed subsets of $I$. A set $X\in{\mathcal S}$ is closed with 
respect to $\cl$ if and only if $X$ is the greatest element of 
$\{G\in {\mathcal S}\mid \cl(G)=\cl(X)\}$, partially ordered by inclusion.
\end{lem}
{\em Proof:} 
Let $X\in{\mathcal S}$ be closed and let $G\in{\mathcal S}$ such that
$\cl(G)=\cl(X)$, then 
\[\textstyle
G\subset\cl(G)=\cl(X)=X.
\]
Conversely, let $X$ be the greatest element of all $G\in{\mathcal S}$ for 
which $\cl(G)=\cl(X)$ holds. Since $\cl(X)\in{\mathcal S}$ and since
$\cl(\cl\,X)=\cl(X)$ we have $X\supset\cl(X)$. As $\cl$ is extensive,
we obtain $X=\cl(X)$ which shows that $X$ is closed.
\hspace*{\fill}$\square$\\
\par
As a standard notation for the article, let $A$ denote a finite-dimensional 
Euclidean vector space, $C\subset A$ a convex subset, $U\subset A$ a 
linear subspace, and 
\[\textstyle
\pi_U:A \to A
\]
the orthogonal projection onto $U$. We frequently write $\pi$ instead of 
$\pi_U$. An {\em exposed face} of $C$ is defined to be either the empty 
set or any subset of $C$ of the form
\begin{equation}\textstyle\label{eq:exp-f}
F_C(a):=\argmin\{\langle x,a\rangle\mid x\in C\},
\qquad
a\in A.
\end{equation}
We denote by $\cE(C)$ the set of exposed faces of $C$. If $x\in C$ and
$\{x\}$ is an exposed face, then $x$ is called an {\em exposed point}.
\begin{defn}[Complete lattice]
Let $\varphi:\cK_1\to\cK_2$ be a map between two partially ordered sets 
$\cK_1,\cK_2$. The map $\varphi$ is {\em isotone} if 
$x\leq y\implies\varphi(x)\leq\varphi(y)$ holds and $\varphi$ is 
{\em antitone} if $x\leq y\implies\varphi(x)\geq\varphi(y)$ holds.
Let $\cK$ be a {\em lattice}, that is a partially ordered set where the 
infimum $k\wedge \ell$ and supremum $k\vee\ell$ of each pair of elements 
$k,\ell\in\cK$ exists. The lattice $\cK$ is {\em complete} if the infimum 
and supremum of an arbitrary subset of $\cK$ exist. For a complete lattice
$\cK$, the infimum of $\emptyset$ is defined to be the greatest element of 
$\cK$ and the supremum of $\emptyset$ is the smallest element.
\end{defn}
\par
The set of exposed faces $\cE(C)$, partially ordered by inclusion, forms a 
complete lattice whose infimum is the intersection 
\cite{Barker1978,LoewyTam1986,Weis2012a}. Consider the set of pre-images 
of exposed faces of $\pi(C)$,
\begin{equation}\textstyle\label{eq:lift-ex}
\cL=\cL_{C,U}=\{\pi|_C^{-1}(F)\mid F\in\cE(\pi(C))\}.
\end{equation}
We denote the smallest element of $\cL$ containing $X\subset C$ by 
\begin{equation}\textstyle\label{eq:clL}
\cl_\cL(X)=\bigcap\{F\in\cL\mid X\subset F\}.
\end{equation}
The set $\cL$, partially ordered by inclusion, is again a complete 
lattice whose infimum is the intersection, for details see
Proposition 5.6 of \cite{Weis2012a}. Hence, the set-valued map 
$\cl_\cL:2^C\to 2^C$ is a closure operation whose closed sets are 
the elements of $\cL$. Clearly $\cL\subset\cE(C)$ holds so 
Lemma~\ref{lem:1} shows the following.
\begin{lem}\label{lem:char-L}
Let $F\in\cE(C)$. We have $F\in\cL$ if and only if $F$ is the 
greatest element of $\{G\in\cE(C)\mid \cl_{\cL}(G)=\cl_{\cL}(F)\}$, 
partially ordered by inclusion.
\end{lem}
\par
Let us reformulate the condition $\cl_{\cL}(G)=\cl_{\cL}(F)$ of 
Lemma~\ref{lem:char-L} in terms of normal cones. First, we recall 
from Proposition 5.6 of \cite{Weis2012a} that the projection
\begin{equation}\textstyle\label{eq:lift-iso}
\pi|_\cL:\cL\to\cE(\pi(C))
\end{equation}
is an isotone lattice isomorphism. Here, the set-valued map 
$2^A \to2^A$ induced by the projection $\pi:A\to A$ is denoted by 
the symbol $\pi$, too. We denote the smallest exposed face of $C$ 
containing $X\subset C$ by 
\[\textstyle
\cl_{\cE(C)}(X)=\bigcap\{F\in\cE(C)\mid X\subset F\}.
\]
\begin{lem}\label{lem:cl-eq}
For all  $X\subset C$ we have 
$\pi(\cl_\cL(X))=\cl_{\cE(\pi(C))}(\pi(X))$.
\end{lem}
{\em Proof:} 
Using the lattice isomorphism of (\ref{eq:lift-iso}), one obtains
from (\ref{eq:clL}) the equation
\[\textstyle
\pi(\cl_\cL(X))=
\bigcap\{F\in\cE(\pi(C))\mid X\subset\pi|_C^{-1}(F)\}.
\]
For $F\in\cE(\pi(C))$ one has $\pi\circ\pi|_C^{-1}(F)=F$. 
Hence $X\subset\pi|_C^{-1}(F)$ implies $\pi(X)\subset F$. 
Conversely, $\pi(X)\subset F$ implies 
\[\textstyle
X\subset\pi|_C^{-1}\circ\pi(X)\subset\pi|_C^{-1}(F)
\]
which completes the proof.
\hspace*{\fill}$\square$\\
\par
A vector $u\in A$ is a {\em normal vector} (inward pointing) of $C$ at 
$x\in C$ if $\langle y-x,u\rangle\geq 0$ for all $y\in C$. The 
{\em normal cone} $N_C(x)$ of $C$ at $x$ is the set of all normal 
vectors of $C$ at $x$. The {\em relative interior} ${\rm ri}(C)$ of 
$C$ is the interior of $C$ in the topology of the affine hull of $C$. 
The {\em normal cone} $N_C(X)$ of $C$ at a non-empty convex subset 
$X\subset C$ is defined to be the normal cone of $N_C(x)$ of $C$ at any
point $x\in{\rm ri}(X)$ (the definition does not depend on $x$, see 
\cite{Schneider2014,Weis2012a}). The normal cone of $C$ at $\emptyset$ 
is defined to be $N_C(\emptyset)=A$. Let 
\[\textstyle
\cN(C):=\{N_C(X)\mid\mbox{$X\subset C$ is convex}\}
\] 
denote the set of normal cones of $C$.
\par
We recall basic properties of normal cones employed later. First, 
$\cN(C)$ is, partially ordered by inclusion, a complete lattice whose 
infimum is the intersection \cite{Weis2012a}. Second, normal cones do
not decrease under the closure operation $\cl_{\cE(C)}$,
\begin{equation}\textstyle\label{eq:nc-of-sup}
N_C(X)=N_C(\cl_{\cE(C)}(X)),
\qquad\mbox{for convex $X\subset C$.}
\end{equation}
The equation (\ref{eq:nc-of-sup}) is proved in Lemma 4.6 of \cite{Weis2012a} 
for faces $X$ of $C$, but is true as stated here%
\footnote{%
Lemma 4.4 of \cite{Weis2012a} and the last statement of Lemma 4.6 of 
\cite{Weis2012a} are formulated for faces, but the proofs are 
straight forward to generalize from faces to arbitrary convex subsets.}
for convex subsets $X\subset C$. Thereby, a {\em face} of $C$ is a 
convex subset of $C$ which contains every closed segment contained in 
$C$ whose open segment it meets. Third, if $C$ is not a singleton, then 
Proposition~4.7 of \cite{Weis2012a} shows that
\begin{equation}\textstyle\label{iso:EN}
N_C:\cE(C)\to\cN(C), 
\quad F\mapsto N_C(F)
\end{equation}
is an antitone lattice isomorphism. Fourth, we interpret normal cones 
of $\pi(C)$ as subsets of $U$. For $x\in\pi(C)$ and $y\in\pi|_C^{-1}(\{x\})$, 
the normal cone of $\pi(C)$ at $x$ is $N_{\pi(C)}(x)=N_C(y)\cap U$, see 
Lemma~5.9 of \cite{Weis2012a}. Hence
\begin{equation}\textstyle\label{eq:nc-proj}
N_{\pi(C)}(\pi(X))=N_C(X)\cap U,
\qquad\mbox{for convex $X\subset C$},
\end{equation}
because $\pi({\rm ri}\,X)\subset{\rm ri}(\pi(X))$, see for example 
\cite{Rockafellar1970}. By convention, $N_{\pi(C)}(\emptyset)=U$. 
\begin{pro}\label{pro:char-L-nc}
Let $\pi(C)$ be not a singleton and let $F\in\cE(C)$. Then $F\in\cL$ 
holds if and only if $F$ is the greatest element of 
$\{G\in\cE(C)\mid N_C(G)\cap U=N_C(F)\cap U\}$,
partially ordered by inclusion.
\end{pro}
{\em Proof:} 
Let $F\in\cE(C)$. Lemma~\ref{lem:char-L} shows that $F$ lies in $\cL$ 
if and only if $F$ is the greatest among all exposed faces $G\in\cE(C)$ 
which have the same closure in $\cL$,
\[\textstyle
\cl_\cL(G)=\cl_\cL(F).
\]
We transform the condition into five equivalent conditions.
\begin{align*}
\pi(\cl_\cL\,G) 
 &=
\pi(\cl_\cL\,F)\\
\cl_{\cE(\pi(C))}\,\pi(G)
 &=
\cl_{\cE(\pi(C))}\,\pi(F)\\
N_{\pi(C)}\left(
\cl_{\cE(\pi(C))}\,\pi(G)
\right)
 &=
N_{\pi(C)}\left(
\cl_{\cE(\pi(C))}\,\pi(F)
\right)\\
N_{\pi(C)}\left( \pi(G) \right)
 &=
N_{\pi(C)}\left( \pi(F) \right)\\
N_C(G)\cap U
 &=
N_C(F)\cap U.
\end{align*}
These conditions follow, respectively, from 
the isomorphisms (\ref{eq:lift-iso}), 
Lemma~\ref{lem:cl-eq}, 
the isomorphism (\ref{iso:EN}), 
equation (\ref{eq:nc-of-sup}), 
and equation (\ref{eq:nc-proj}).
\hspace*{\fill}$\square$\\
%
%
%%%%%%%%%%%%%%%%%%%%%%%%%%%%%%%%%%%%%%%%%%%%%%%%%%%%%%%%%%%%%%%%%%%%%%%%%%%%
%%%%%%%%%%%%%%%%%%%%%%%%%%%%%%%%%%%%%%%%%%%%%%%%%%%%%%%%%%%%%%%%%%%%%%%%%%%%
%%%%%%%%%%%%%%%%%%%%%%%%%%%%%%%%%%%%%%%%%%%%%%%%%%%%%%%%%%%%%%%%%%%%%%%%%%%%
%%%%%%%%%%%%%%%%%%%%%%%%%%%%%%%%%%%%%%%%%%%%%%%%%%%%%%%%%%%%%%%%%%%%%%%%%%%%
%%%%%%%%%%%%%%%%%%%%%%%%%%%%%%%%%%%%%%%%%%%%%%%%%%%%%%%%%%%%%%%%%%%%%%%%%%%%
%
\section{Geometric representations of projections}
\label{sec:lattices-projections}
\par
Classical statistics and the statistics of quantum mechanics are 
special among more general statistical theories 
\cite{Holevo2012,Gross-etal2010} because they can be described in 
the setting of C*-algebras. We recall some lattice isomorphisms 
in the C*-algebraic context \cite{AlfsenShultz2001}.
\par
Here and subsequently, let $\cA\subset M_n$ be a complex *-subalgebra 
including the $n$-by-$n$ identity matrix $\id$, and let $A\subset\cA$ 
denote the real subspace of hermitian matrices. The algebra $\cA$ 
is partially ordered by the relation $a\preceq b$, $a,b\in\cA$, also 
denoted $b\succeq a$, which means that $b-a$ is positive semi-definite. 
This partial ordering is known as the {\em L\"owner partial ordering}. 
The fundamental events of the statistical theory form the set of 
{\em projections} of $\cA$,
\[\textstyle
\cP_\cA:=\{p\in\cA\mid p=p^*=p^2\}.
\]
The set $\cP_\cA$, endowed with the L\"owner partial ordering, is a complete 
lattice \cite{AlfsenShultz2001} and
\begin{equation}\textstyle\label{eq:iso-im}
\cP_\cA\to\{{\rm image}(p)\mid p\in\cP_\cA\},
\quad
p\mapsto{\rm image}(p)\subset\bC^n
\end{equation}
is a lattice isomorphism, where subspaces of $\bC^n$ are partially ordered by 
inclusion. The infimum of two subspaces is their intersection.
\par
In quantum mechanics, a hermitian matrix has the interpretation of 
energy operator ({\em Hamiltonian}) and its eigenvalues are the 
possible energy values \cite{BengtssonZyczkowski2017}. We denote 
the real vector space of hermitian matrices by
\[\textstyle
A:=\{a\in\cA\mid a^*=a\}.
\] 
The smallest eigenvalue $\lambda_0(a)$ of $a\in A$ is the {\em ground energy} 
of $a$, the corresponding eigenspace of $a$ is the {\em ground space} of $a$, 
and the projection $p_0(a)\in\cP_\cA$ onto the ground space is the 
{\em ground projection} of $a$. This defines a map
\begin{equation}\textstyle\label{def:p0}
p_0:A\to \cP_\cA.
\end{equation}
Consider the {\em state space} 
$C_\cA=\{\rho\in\cA\mid\rho\succeq 0,\tr(\rho)=1\}$ 
of positive semi-definite trace-one matrices in $\cA$. If 
$\rho\in C_\cA$, then the projection onto ${\rm image}(\rho)$ 
is called {\em support projection} of $\rho$ and will be 
denoted by $\supp(\rho)\in\cP_\cA$. We call $\rho$ a 
{\em ground state} of $a\in A$ if $\supp(\rho)\preceq p_0(a)$. 
\par
The commutative algebra $\cA\subset M_n$ of diagonal matrices 
may be seen as the space of functions 
\begin{equation}\textstyle\label{eg:classical}
\cA\cong\bC^X\cong\{f:X\to\bC\}
\end{equation}
on a {\em configuration space} $X$ of cardinality $|X|=n$. A basis 
of $\bC^X$ is given by $(\delta_x)_{x\in X}$,
\begin{equation}\label{eq:delta}\textstyle
\delta_x(y)
=\left\{\begin{array}{ll} 
0 & \mbox{if $y\neq x$}\\
1 & \mbox{if $y=x$}
\end{array}\right.,
\qquad
x,y\in X.
\end{equation}
Hermitian matrices are in one-to-one correspondence with real functions 
$A=\bR^X$. The state space $C_{\cA}$ of $\bC^X$ is the simplex of 
probability distributions on $X$,
\[\textstyle
\Delta_X:=\{f:X\to\bR \mid \forall x\in X:f(x)\geq 0
\mbox{ and } \sum_{x\in X}f(x)=1\}.
\]
The projections are in one-to-one correspondence with $0$-$1$-functions 
$\cP_\cA\cong\{0,1\}^X$, which we identify with the power set $2^X$ of 
subsets of $X$. The lattice $\cP_\cA$ is a Boolean algebra 
\cite{Aigner1997,Birkhoff1967}, partially ordered by inclusion. For
$p,q\in\cP_\cA$, the complementary projection $p'=\id-p$ of $p$ is the 
set-theoretical complement $p'=X\setminus p$, the infimum of $p$ and $q$ 
is the intersection $p\wedge q=p\cap q$ and the supremum is the union 
$p\vee q=p\cup q$. While ground spaces are vector spaces, it should be 
save to call the finite set
\[\textstyle
p_0(f)=\argmin\{f(x)\mid x\in X\}\subset X
\]
{\em ground space} of $f\in A$. We refer to $x\in p_0(f)$ as a
{\em ground configuration} of $f$. 
\par
We return to an arbitrary *-subalgebra $\cA\subset M_n$ with $\id\in\cA$.
Opposed to the simplex, $C_\cA$ has no intuitive visualization in the 
three-dimensional space \cite{Bengtsson-etal2013}. The geometry of $C_\cA$
is easily grasped by the lattice isomorphism 
\begin{equation}\textstyle\label{eq:PE}
\phi=\phi_\cA:\cP_\cA\to\cE(C_\cA),
\quad
p\mapsto\{\rho\in C_\cA\mid\supp(\rho)\preceq p\},
\end{equation}
from projections to exposed faces, see for example Corollary~3.36 of 
\cite{AlfsenShultz2001}. With notation from (\ref{eq:exp-f}) and 
(\ref{def:p0}), the projections may be taken to be ground projections, 
we have 
\begin{equation}\textstyle\label{eq:F=PhiP}
F_{C}=\phi_\cA\circ p_0.
\end{equation}
To study normal cones we recall that the relative interior of $F_C(a)$ is 
\begin{equation}\textstyle\label{eq:riF}
{\rm ri}\,F_{C}(a)=\{\rho\in C\mid\supp(\rho)= p_0(a)\},
\qquad a\in A.
\end{equation}
See Proposition 2.9 of \cite{Weis2011} for a proof of (\ref{eq:F=PhiP})
and (\ref{eq:riF}).
\par
Like exposed faces, the normal cones of $C_\cA$ have a simple 
algebraic representation. Proposition 2.11 of \cite{Weis2011} shows 
\[\textstyle
N_C(\rho)=\{a\in A\mid \supp(\rho)\preceq p_0(a)\},
\qquad 
\rho\in C_\cA.
\]
The normal cone at a convex subset being equal to the normal cone at 
any relative interior point of the convex set, the preceding equation 
and (\ref{eq:riF}) show
\begin{equation}\textstyle\label{eq:NFace}
N_C(F_C(a))
=\{b\in A\mid p_0(a)\preceq p_0(b)\},
\qquad 
a\in A.
\end{equation}
We observe that $C_\cA$ is not a singleton if $\cA\not\cong\bC$.
In that case we obtain, combining (\ref{eq:PE}) and (\ref{iso:EN}), 
the antitone lattice isomorphism 
\begin{equation}\textstyle\label{eq:P-N}
\nu:\cP_\cA\to\cN_{C_\cA},
\quad
p\mapsto N_{C_\cA}\circ\phi_\cA(p).
\end{equation}
Notice that, since $p_0:A\to\cP_\cA$ is onto $\cP_\cA\setminus\{0\}$, 
the equations (\ref{eq:F=PhiP}) and (\ref{eq:NFace}) show
\begin{equation}\textstyle\label{eq:nu-formula}
\nu(p)
=\{a\in A\mid p\preceq p_0(a)\},
\qquad p\in\cP_\cA.
\end{equation}
%
%%%%%%%%%%%%%%%%%%%%%%%%%%%%%%%%%%%%%%%%%%%%%%%%%%%%%%%%%%%%%%%%%%%%%%%%%%%%
%%%%%%%%%%%%%%%%%%%%%%%%%%%%%%%%%%%%%%%%%%%%%%%%%%%%%%%%%%%%%%%%%%%%%%%%%%%%
%%%%%%%%%%%%%%%%%%%%%%%%%%%%%%%%%%%%%%%%%%%%%%%%%%%%%%%%%%%%%%%%%%%%%%%%%%%%
%%%%%%%%%%%%%%%%%%%%%%%%%%%%%%%%%%%%%%%%%%%%%%%%%%%%%%%%%%%%%%%%%%%%%%%%%%%%
%%%%%%%%%%%%%%%%%%%%%%%%%%%%%%%%%%%%%%%%%%%%%%%%%%%%%%%%%%%%%%%%%%%%%%%%%%%%
%
\section{Geometric representations of ground projections}
\label{sec:lattices}
\par
We discuss the lattice of ground projections of a linear subspace 
of hermitian matrices. We recall that the lattice is isomorphic to 
the lattices of exposed faces and normal cones of a linear image
of the state space. We point out the fundamental property that the
lattice of ground projections of a linear subspace is coatomistic.
\par
Here and subsequently, let $U\subset A$ a linear subspace and 
$\pi:A \to A$ the orthogonal projection onto $U$. 
The isotone lattice isomorphism (\ref{eq:PE}) restricts to another useful 
map. Let
\[\textstyle
\cP(U):=\{p_0(u)\mid u\in U\}\cup\{0\}
\]
denote the set of ground projections of $U$.
The identity (\ref{eq:F=PhiP}) shows that $\phi$ restricts to the bijection 
\begin{equation}\textstyle\label{eq:iso-PU-L}
\phi|_{\cP(U)}:\cP(U)\to\cL
\end{equation}
onto the lattice of pre-images of exposed faces of $\pi(C_\cA)$, introduced 
in (\ref{eq:lift-ex}). Since $\cL$ is a complete lattice whose infimum is
the intersection, it follows that $\phi|_{\cP(U)}$ is an isotone lattice 
isomorphism and that $\cP(U)$, partially ordered by the L\"owner partial 
ordering, is a complete lattice whose infimum is given in terms of 
intersection of images (\ref{eq:iso-im}). In the classical case
(\ref{eg:classical}) of $\cA\cong\bC^X$, the infimum in $\cP(U)$ is simply 
the intersection of subsets of $X$. We summarize.
\begin{lem}\label{lem:PU}
Let $U\subset A$ be a linear subspace. Then $\cP(U)$ is a complete lattice. 
The infimum of $p,q\in\cP(U)$ is the projection $r\in\cP_\cA$ whose image 
is the intersection ${\rm image}(r)={\rm image}(p)\cap{\rm image}(q)$. In 
the classical case (\ref{eg:classical}) we have $p\wedge q=p\cap q$.
\end{lem}
\par
There are two important lattice isomorphisms connecting $\cP(U)$ to
the projection $\pi(C_\cA)$ of the state space $C_\cA$ onto $U$.
The isomorphisms (\ref{eq:iso-PU-L}) and (\ref{eq:lift-iso}) show that 
\begin{equation}\textstyle\label{eq:PEpi}
\pi\circ\phi|_{\cP(U)}:\cP(U)\to\cE(\pi(C))
\end{equation} 
is an isotone lattice isomorphism \cite{Weis2018}. 
If $\pi(C)$ is not a singleton, then by (\ref{iso:EN}) 
\begin{equation}\textstyle\label{eq:PNpi}
N_{\pi(C)}\circ\pi\circ\phi|_{\cP(U)}:\cP(U)\to\cN(\pi(C))
\end{equation} 
is an antitone lattice isomorphism. Using (\ref{eq:nc-proj}) and the 
map $\nu$ from (\ref{eq:P-N}), we notice 
\begin{equation}\textstyle\label{eq:NU0}
N_{\pi(C)}\circ\pi\circ\phi(p)=\nu(p)\cap U,
\qquad
p\in\cP_\cA,
\end{equation} 
where $\nu(p)$ is a normal cone of the state space $C_\cA$.  
\begin{rem}
The convex set $\pi(C_\cA)$ is affinely isomorphic to convex sets 
studied in other fields: 
{\em Algebraic polar} of a spectrahedron \cite{RamanaGoldman1995}, 
{\em joint algebraic numerical range} \cite{Mueller2010}, 
and {\em state space} of an operator system \cite{Paulsen2002}. 
See \cite{Weis2018} for details about the latter.
\end{rem}
\par
The lattice of exposed faces $\cE(\pi(C_\cA))$ is generated by 
certain maximal elements.
\begin{defn}[Atoms and coatoms]
Let $\cK$ be a complete lattice with smallest element~$0$ and greatest 
element~$1$. We say $0$ and $1$ are {\em improper} elements of $\cK$
and we call $x\in\cK$ a {\em proper} element of $\cK$ if and only 
$x\not\in\{0,1\}$. An {\em atom} of $\cK$ is a minimal element of 
$\cK\setminus\{0\}$. The lattice $\cK$ is {\em atomistic} if each element 
is the supremum of the atoms which it contains. 
A {\em coatom} of $\cK$ is a maximal element of 
$\cK\setminus\{1\}$. The lattice $\cK$ is {\em coatomistic} if each 
element is the infimum of the coatoms in which it is contained.
\end{defn}
\par
The lattice $\cE(\pi(C_\cA))$ is coatomistic \cite{Weis2012a} and
so is $\cP(U)$ by virtue of (\ref{eq:PEpi}), as was shown in 
\cite{Weis2018}.
\begin{lem}\label{lem:coatomistic}
Let $U\subset A$ be a linear subspace. Then the lattice $\cP(U)$ is 
coatomistic.
\end{lem}
\begin{figure}[t]
\includegraphics[height=3cm]{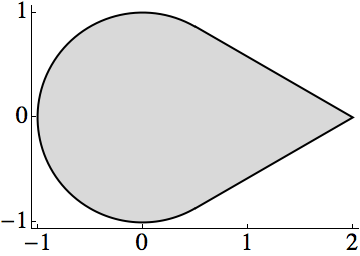}
\caption{\label{fig:drop}
The convex hull of the unit disk and the point $(2,0)$ has two 
boundary segments containing only one atom, $\{(2,0)\}$, of the 
lattice of exposed faces. They cannot be written as 
suprema of atoms.}
\end{figure}
\par
The lattice $\cE(\pi(C))$ is generally not atomistic. An example 
with $n=3$ is depicted in Figure~\ref{fig:drop} in coordinates 
$\{(\langle \rho,a_1\rangle,\langle \rho,a_2\rangle)\mid\rho\in C_{M_3}\}\subset\bR^2$
where $U$ is the span of
\begin{equation}\label{eq:example}\textstyle
a_1=
\left(\begin{smallmatrix}0 & 1 & 0\\1 & 0 & 0\\
0 & 0 & 2\end{smallmatrix}\right)
\quad\mbox{and}\quad
a_2=
\left(\begin{smallmatrix}0 & -\ii & 0\\\ii & 0 & 0\\
0 & 0 & 0\end{smallmatrix}\right).
\end{equation}
In the classical case (\ref{eg:classical}) the state space 
$C_\cA=\Delta_X$ is a simplex, $\pi(\Delta_X)$ a polytope, and
$\cE(\pi(\Delta_X))$ atomistic\footnote{%
The term (co-) atomistic employed here is 
called (co-) atomic in \cite{Ziegler1995}.}.
The atoms are the exposed points and the coatoms the {\em facets} of
$\pi(\Delta_X)$, that is exposed faces of codimension one, see Theorem~2.7 
of \cite{Ziegler1995}.
\begin{rem}[Top to bottom {\em via} minors]%
\label{rem:top-to-bottom-minors}
If $\id\in U$ then every non-zero element of $\cP(U)$ is the ground 
projection of a positive semi-definite $u\in U$ with $p_0(u)=\ker(u)$.
The Sylvester criterion \cite{HornJohnson2012} shows for $k=1,\ldots,n$ 
that $\{p\in\cP(U)\mid{\rm rk}(p)\geq k\}$ is formed by the ground 
projections of solutions of the real algebraic system 
\\[.5\baselineskip]
\centerline{\begin{tabular}{lll}
$\{ u\in U \,|$ & \mbox{the minors of $u$ of size $n-k+1$ are zero and} \\
              & \mbox{the principal minors of $u$ of size at most $n-k$ 
  are non-negative} $\}$.
\end{tabular}}\\[.5\baselineskip]
Lemma~\ref{lem:coatomistic} suggests to solve the algebraic systems by
decreasing order of the rank $k=n-1,n-2,\ldots,1$. This order has also 
the advantage to start with the system of the smallest algebraic degree.
\end{rem}
%
%%%%%%%%%%%%%%%%%%%%%%%%%%%%%%%%%%%%%%%%%%%%%%%%%%%%%%%%%%%%%%%%%%%%%%%%%%%%
%%%%%%%%%%%%%%%%%%%%%%%%%%%%%%%%%%%%%%%%%%%%%%%%%%%%%%%%%%%%%%%%%%%%%%%%%%%%
%%%%%%%%%%%%%%%%%%%%%%%%%%%%%%%%%%%%%%%%%%%%%%%%%%%%%%%%%%%%%%%%%%%%%%%%%%%%
%%%%%%%%%%%%%%%%%%%%%%%%%%%%%%%%%%%%%%%%%%%%%%%%%%%%%%%%%%%%%%%%%%%%%%%%%%%%
%%%%%%%%%%%%%%%%%%%%%%%%%%%%%%%%%%%%%%%%%%%%%%%%%%%%%%%%%%%%%%%%%%%%%%%%%%%%
%
\section{A variation principle for ground spaces}
\label{sec:var-gs}
\par
The variation principle of Proposition~\ref{pro:char-L-nc} is applied here to 
ground projections.
\begin{lem}\label{lem:char-PU-nu}
Let $U\subset A$ be a linear subspace such that $\pi(C)$ is no singleton, 
and let $p\in\cP_\cA$. Then $p\in\cP(U)$ holds if and only if $p$ is the 
greatest element of $\{q\in\cP_\cA\mid\nu(q)\cap U=\nu(p)\cap U\}$ in the 
L\"owner partial ordering.
\end{lem}
{\em Proof:} 
Proposition~\ref{pro:char-L-nc} shows for $F\in\cE(C)$ that $F\in\cL$ 
if and only if $F$ is the greatest element of 
\[\textstyle
\{G\in\cE(C)\mid N_C(G)\cap U=N_C(F)\cap U\}.
\]
The lattice isomorphisms $\phi:\cP_\cA\to\cE(C)$ from (\ref{eq:PE}) and 
$\phi|_{\cP(U)}:\cP(U)\to\cL$ from (\ref{eq:iso-PU-L}) show for 
$p\in\cP_\cA$ that $p\in\cP(U)$ if and only if $p$ is the greatest element 
of 
\[\textstyle
\{q\in\cP_\cA\mid N_C\circ\phi(q)\cap U=N_C\circ\phi(p)\cap U\}.
\]
The definition (\ref{eq:P-N}) of $\nu$ finishes the proof.
\hspace*{\fill}$\square$\\
\par
To employ Lemma~\ref{lem:char-PU-nu} algebraically, we use
(\ref{eq:nu-formula}) and write
\begin{equation}\label{eq:nc-alg}\textstyle
\nu(p)
=\{a\in A\mid p\preceq p_0(a)\}
=\bR\id + p'\cA^+p',
\qquad p\in\cP_\cA.
\end{equation}
Recall from (\ref{eq:NU0}) that $\nu(p)\cap U=(\bR\id + p'\cA^+p')\cap U$ 
is a normal cone of $\pi(C)$.
\begin{defn}\label{def:K}
For all $p\in\cP_\cA$, let $K(p):=p'\cA^+p'\cap U$.
\end{defn}
\par
The substitution of $\nu(p)\cap U$ with $K(p)$ is not always faithful.
For example, let $\cA=M_2$, let $U$ be the span of 
$X=\left(\begin{smallmatrix}0 & 1\\1 & 0\end{smallmatrix}\right)$ and
$Y=\left(\begin{smallmatrix}0 & -\ii\\\ii & 0\end{smallmatrix}\right)$,
and let $u\in U$ have spectral norm one. Then $p_0(u)=\tfrac{1}{2}(\id-u)$
is the ground projection of $u$ and 
\[\textstyle
\nu\left(p_0(u)\right)\cap U
=\left\{\lambda\cdot u \mid \lambda\geq 0\right\},
\]
while $K(p)=\{0\}$ for all $p\in\cP_\cA$. 
\begin{lem}\label{lem:Keqnu}
Let $U\subset A$ be a linear subspace with $U\supsetneq\bR\id$. 
Then for all $p,q\in\cP_\cA$ we have 
$\nu(p)\cap U=\nu(q)\cap U\iff K(p)=K(q)$.
\end{lem}
{\em Proof:} 
We prove ``$\Leftarrow$''. We have for all 
$p\in\cP_\cA$
\begin{equation}\textstyle\label{eq:nufromK}
\nu(p) \cap U 
= (\bR\id + p'\cA^+p') \cap U 
= \bR\id + (p'\cA^+p' \cap U) 
= \bR\id+K(p).
\end{equation}
The first equality of (\ref{eq:nufromK}) holds by (\ref{eq:nc-alg}), 
the second equality requires $\id\in U$, see equation (22) of 
\cite{Weis2012a}. 
We prove ``$\Rightarrow$''. Unless stated otherwise, $\id\in U$ is 
not assumed in the sequel. For $p\in\cP_\cA$ we have
\[\textstyle
K(p) = \{u\in U \mid u\succeq 0, pu=0 \}.
\]
Let $p\neq 0$. Then for any $a\in A$ such that $pa=0$, each of the 
statements $a\succeq 0$ or $p\preceq p_0(a)$ implies $\ker(a)=p_0(a)$. 
Hence
\[\textstyle
K(p) = \{u\in U \mid p\preceq p_0(u),pu=0 \}, 
\qquad p\neq 0,
\]
that is by (\ref{eq:nc-alg}) 
\begin{equation}\label{eq:lin-sec}\textstyle
K(p) = \{u\in \nu(p)\cap U \mid pu=0 \}, 
\qquad p\neq 0.
\end{equation}
To finish the proof it suffices to show that $\nu(0)\cap U\neq \nu(p)\cap U$. 
We have
\[\textstyle
 \nu(0)\cap U
 = N_C(\emptyset)\cap U
 = A \cap U
 = U
 = N_{\pi(C)}(\emptyset),
\] 
and by (\ref{eq:NU0}) and (\ref{eq:nc-of-sup})
\[\textstyle
\nu(p)\cap U
 = N_{\pi(C)}(\pi\circ\phi(p))
 = N_{\pi(C)}(\cl_{\cE(\pi(C))}(\pi\circ\phi(p))
\] 
holds. Since $p\neq 0$, the exposed face $\cl_{\cE(\pi(C))}(\pi\circ\phi(p))$ 
is non-empty. Assuming $\pi(C)$ is not a singleton, which is 
implied by $U\supsetneq\bR\id$, the map 
$N_{\pi(C)}:\cE(\pi(C))\to\cN(\pi(C))$ is the isomorphism 
(\ref{iso:EN}). This completes the proof.
\hspace*{\fill}$\square$\\
\par
Notice that the statement of Lemma~\ref{lem:Keqnu} is wrong for 
$U=\bR\id$. Here $\nu(p)\cap U=U$ for all $p\in\cP_\cA$, while 
$K(0)=\{\lambda\id\mid\lambda\geq 0\}$ and 
$K(p)=\{0\}$ for all $p\in\cP_\cA\setminus\{0\}$.
\begin{thm}\label{thm:char-PU-K}
Let $U\subset A$ be a linear subspace with $\id\in U$ and let 
$p\in\cP_\cA$. Then $p\in\cP(U)$ holds if and only if $p$ is the 
greatest element of $\{q\in\cP_\cA\mid K(q)=K(p)\}$ in the 
L\"owner partial ordering.
\end{thm}
{\em Proof:} 
The case $U=\bR\id$ is verified in the preceding paragraph. If 
$U\supsetneq\bR\id$ then the claim follows from 
Lemmas~\ref{lem:char-PU-nu} and~\ref{lem:Keqnu}.
\hspace*{\fill}$\square$\\
\par
Using the linear spaces $L(p):=p'Ap'\cap U$, $p\in\cP_\cA$, may simplify matters.
\begin{lem}\label{lem:linear}
Let $U\subset A$ be a linear subspace, let $p\in\cP_\cA$.
Then $K(p)=L(p)\cap\cA^+$. In the classical case (\ref{eg:classical})
we have $L(p)=(pAp+U^\perp)^\perp$.
\end{lem}
%
%
%
%%%%%%%%%%%%%%%%%%%%%%%%%%%%%%%%%%%%%%%%%%%%%%%%%%%%%%%%%%%%%%%%%%%%%%%%%%%%
%%%%%%%%%%%%%%%%%%%%%%%%%%%%%%%%%%%%%%%%%%%%%%%%%%%%%%%%%%%%%%%%%%%%%%%%%%%%
%%%%%%%%%%%%%%%%%%%%%%%%%%%%%%%%%%%%%%%%%%%%%%%%%%%%%%%%%%%%%%%%%%%%%%%%%%%%
%%%%%%%%%%%%%%%%%%%%%%%%%%%%%%%%%%%%%%%%%%%%%%%%%%%%%%%%%%%%%%%%%%%%%%%%%%%%
%%%%%%%%%%%%%%%%%%%%%%%%%%%%%%%%%%%%%%%%%%%%%%%%%%%%%%%%%%%%%%%%%%%%%%%%%%%%
%
\section{Coatoms of the lattice of ground spaces}
\label{sec:coatoms-gs}
\par
Here we characterize the coatoms of the lattice of 
ground projections $\cP(U)$.
\begin{thm}\label{thm:ray-coatoms}
Let $U\subset A$ be a linear subspace with $\id\in U$ and let $p\in\cP(U)$. 
\begin{enumerate}
\item 
The projection $p$ is a coatom of $\cP(U)$ if and only if $K(p)$ is a ray. 
\item
Let $p\neq 0$ and $d=\dim\,K(p)$. There are coatoms $q_1,\ldots,q_d$ of 
$\cP(U)$ such that $p=q_1\wedge\ldots\wedge q_d$ and such that the rays 
$K(q_1),\ldots,K(q_d)$ are linearly independent exposed extreme rays of 
$K(p)$. 
\end{enumerate}
\end{thm}
{\em Proof:}
Let $U=\bR\id$. Then $\cP(U)=\{0,\id\}$ and indeed 
$K(0)=\{\lambda\id\mid\lambda\geq 0\}$ is a ray while $K(\id)=\{0\}$. The 
second assertion is true as the infimum of $\emptyset$ is $\id$.
\par
In the following we assume $U\supsetneq\bR\id$. Let $U_0\subset U$ be the 
space of traceless matrices in $U$ and notice $\cP(U_0)=\cP(U)$. The convex 
set $\pi_{U_0}(C)$ is a {\em proper convex set} in the sense of 
\cite{Weis2018}, that is $\pi_{U_0}(C)$ has non-empty interior in $U_0$ 
and $\cE(\pi_{U_0}(C))$ contains a proper exposed face. Under these 
(technical) assumptions, Theorem~6.2 and Corollary~6.3 of \cite{Weis2018} 
show that any non-empty face of any normal cone $N$ of $\pi_{U_0}(C)$ lies 
in $\cN_{\pi_{U_0}(C)}$. Hence, Remark~2.2(4) of \cite{Weis2018} shows
that $N$ has $\dim(N)$ linearly independent exposed rays which lie in 
$\cN_{\pi_{U_0}(C)}$. These properties of $\pi_{U_0}(C)$ are used in the 
sequel without further mention.
\par
Since $\nu(p)\pm\bR\id\subset\nu(p)$ holds, we have 
$\bR\id + ( U_0 \cap \nu(p) ) = (\bR\id + U_0 ) \cap \nu(p)$ by 
Lemma~5.1 of \cite{Weis2012a}. Hence
\begin{equation}\textstyle\label{eq:NUU0}
\nu(p)\cap U
= \bR\id + (\nu(p) \cap U_0),
\qquad p\in\cP_\cA.
\end{equation}
Notice that the right-hand side of (\ref{eq:NUU0}) is the direct
sum of a line and a convex cone, the cone being pointed if $p\neq 0$. 
By (\ref{eq:nufromK}) we have 
\begin{equation}\textstyle\label{eq:recnuK}
\nu(p)\cap U=\bR\id+K(p),
\qquad p\in\cP_\cA.
\end{equation}
Notice that the right-hand side of (\ref{eq:recnuK}) is the sum 
of a line and a pointed convex cone, the sum being direct if $p\neq 0$.
\par
{\em Proof of (1).}
The antitone lattice isomorphism $\cP(U_0)\to\cN(\pi_{U_0}(C))$ defined
in (\ref{eq:PNpi}) is the function $p\mapsto\nu(p)\cap U_0$, as observed 
in (\ref{eq:NU0}). Hence, a projection $p\in\cP(U_0)$ is a coatom of 
$\cP(U_0)$ if and only if $\nu(p)\cap U_0$ is an atom of 
$\cN(\pi_{U_0}(C))$. Theorem~3.2 of \cite{Weis2018} shows that the atoms 
of $\cN(\pi_{U_0}(C))$ are the rays in $\cN(\pi_{U_0}(C))$. If 
$p\neq 0$ then by (\ref{eq:NUU0}) and (\ref{eq:recnuK}) the cone 
$\nu(p)\cap U_0$ is a ray if and only if $K(p)$ is a ray. This proves the 
claim for $p\neq 0$. The cone $K(0)$ is not a ray, as $U\supsetneq\bR\id$
and as $\id$ is an interior point of $\cA^+$. In agreement with that, $0$ 
is not a coatom of $\cP(U_0)$, because $\cP(U_0)$, has a proper element by 
(\ref{eq:PEpi}), as $\pi_{U_0}(C)$ has a proper exposed face. 
\par
{\em Proof of (2).}
Let $F$ be a proper exposed face of $\pi_{U_0}(C)$ and 
$d=\dim(N_{\pi_{U_0}(C)}(F))$. Corollary~2.3 of \cite{Weis2018} shows 
that there are coatoms $F_1,\ldots,F_d$ of $\cE(\pi_{U_0}(C))$ such that 
$F=F_1\cap\ldots\cap F_d$ and such that the rays $N_{\pi_{U_0}(C)}(F_i)$, 
$i=1,\ldots,d$, are linearly independent exposed extreme rays of 
$N_{\pi_{U_0}(C)}(F)$. An analogous statement concerning proper ground 
projections $p$ of $U_0$ follows from the lattice isomorphism 
(\ref{eq:PEpi}),
\[\textstyle
\pi_{U_0}\circ\phi|_{\cP(U_0)}:\cP(U_0)\to\cE(\pi_{U_0}(C)).
\]
Let $N=N_{\pi_{U_0}(C)}(\pi_{U_0}\circ\phi(p))$ and $d=\dim(N)$. There are 
coatoms $q_1,\ldots,q_d$ of $\cP(U_0)$ such that $p=q_1\wedge\ldots\wedge q_d$ 
and such that the rays 
$r_i=N_{\pi_{U_0}(C)}(\pi_{U_0}\circ\phi(q_i))$, $i=1,\ldots,d$, are linearly 
independent exposed extreme rays of $N$. Using (\ref{eq:NU0}), we get 
$N=\nu(p)\cap U_0$ and $r_i=\nu(q_i)\cap U_0$, $i=1,\ldots,d$. From 
(\ref{eq:NUU0}) and (\ref{eq:recnuK}) follows that 
$d=\dim(\nu(p)\cap U_0)=\dim\,K(p)$. To see that the rays $K(q_i)$, 
$i=1,\ldots,d$, are exposed rays of $K(p)$, we notice from (\ref{eq:NUU0}) that 
$r_i=\nu(q_i)\cap U_0$ defines an exposed half-plane 
$\bR\id+r_i=\nu(q_i)\cap U$ of $\bR\id+N=\nu(p)\cap U$. The linear space 
$V=\{a\in A \mid pa=0 \}$ intersects $\nu(p)\cap U$ transversally to its 
lineality space $\bR\id$ and the intersection is the pointed convex cone $K(p)$ 
by (\ref{eq:lin-sec}). Hence $(\bR\id+r_i)\cap V$ is an exposed ray of $K(p)$.
Since $q_i\succeq p$ we have $(\bR\id+r_i)\cap V=K(q_i)$. Adding and 
subtracting lineality also shows that the linear independence of 
$\{r_i\}_{i=1}^d$ implies that of $\{K(q_i)\}_{i=1}^d$. This proves the claim 
for proper $p$. The assertion is trivial for $p=\id$.
\hspace*{\fill}$\square$\\
%
%%%%%%%%%%%%%%%%%%%%%%%%%%%%%%%%%%%%%%%%%%%%%%%%%%%%%%%%%%%%%%%%%%%%%%%%%%%%
%%%%%%%%%%%%%%%%%%%%%%%%%%%%%%%%%%%%%%%%%%%%%%%%%%%%%%%%%%%%%%%%%%%%%%%%%%%%
%%%%%%%%%%%%%%%%%%%%%%%%%%%%%%%%%%%%%%%%%%%%%%%%%%%%%%%%%%%%%%%%%%%%%%%%%%%%
%%%%%%%%%%%%%%%%%%%%%%%%%%%%%%%%%%%%%%%%%%%%%%%%%%%%%%%%%%%%%%%%%%%%%%%%%%%%
%%%%%%%%%%%%%%%%%%%%%%%%%%%%%%%%%%%%%%%%%%%%%%%%%%%%%%%%%%%%%%%%%%%%%%%%%%%%
%
\section{A simple non-commutative example}
\label{sec:simple-non-commutative}
\par
We compute ground spaces of the example (\ref{eq:example}) from 
Section~\ref{sec:lattices} and compare with results concerning coatoms
from Section~\ref{sec:coatoms-gs}. 
\par
Let
$\sigma_X=\left(\begin{smallmatrix}0 & 1\\1 & 0\end{smallmatrix}\right)$
and 
$\sigma_Y=\left(\begin{smallmatrix}0 & -\ii\\\ii & 0\end{smallmatrix}\right)$.
The space $U$ is the real span of 
\[
\id, 
\qquad
\left(\begin{smallmatrix}0 & 1 & 0\\1 & 0 & 0\\
0 & 0 & 2\end{smallmatrix}\right)
=
\sigma_X\oplus 2,
\qquad
\left(\begin{smallmatrix}0 & -\ii & 0\\\ii & 0 & 0\\
0 & 0 & 0\end{smallmatrix}\right)
=
\sigma_Y\oplus 0
\qquad
\in M_3.
\]
We write $u\in U$ in the form
\[
u = \lambda\id+\Re(z)(\sigma_X\oplus 2)+\Im(z)\sigma_Y\oplus 0,
\qquad \lambda\in\bR, z\in\bC.
\]
The ground space of $u$ is independent of $\lambda$ and invariant under 
scaling of $u$ with positive scalars. We assume $|z|=1$. Using rank-one 
projections $p(z):=\tfrac{1}{2}
\left(\begin{smallmatrix}1 & \bar z \\ z & 1\end{smallmatrix}\right)$,
\begin{equation}\label{eq:spec-example}\textstyle
u = [(\lambda-1)p(-z) + (\lambda+1)p(z)]\oplus[\lambda + 2\Re(z)]
\end{equation}
is the spectral decomposition of $u$, 
the eigenvalues being $\{\lambda\pm1,\lambda + 2\Re(z)\}$. 
\par
Case a), $\Re(z)=-1/2$. Let $z_\pm:=-\tfrac{1}{2}\pm\ii\tfrac{\sqrt{3}}{2}$ 
and $u_\pm:=2 p(z_\pm)\oplus 0$. For $\lambda=1$ and $z=z_\pm$ we obtain 
$u=u_\pm$ in equation (\ref{eq:spec-example}), and $p_0(u)$ is the rank-two
coatom 
\[
p_\pm:=p_0(u_\pm)=p(-z_\pm)\oplus 1
\]
of $\cP(U)$. The cone
\[
K(p_\pm)=\{u\in U \mid u\succeq 0, p(-z_\pm)\oplus 1\preceq \ker(u) \}
=\{\eta\cdot u_\pm \mid \eta\geq 0\}
\]
is a ray in accord with Theorem~\ref{thm:ray-coatoms}(1).
\par
Case b), $\Re(z)>-1/2$. Taking $\lambda=1$, we have 
$u = 2 p(z)\oplus(1 + 2\Re(z))$ in equation (\ref{eq:spec-example}), 
and $p_0(u)=p(-z)\oplus 0$. Since $p_0(u)\not\preceq p_\pm$ and since 
$p_\pm$ are the only rank-two elements of $\cP(U)$, the projection 
$p_0(u)$ is a coatom of $\cP(U)$. The cone 
\begin{align*}
K(p_0(u)) 
&= \{ u\in U \mid u\succeq 0, p(-z)\oplus 0 \preceq \ker(u) \}\\
&= \{ \eta\cdot(2 p(z) \oplus(1+\Re\,z)) \mid \eta\geq 0 \}
\end{align*}
is a ray in agreement with Theorem~\ref{thm:ray-coatoms}(1).
\par
Case c), $\Re(z)<-1/2$. Equation (\ref{eq:spec-example}) shows 
$p_0(u)=0\oplus 1=p_+\wedge p_-$. The cone 
\begin{align*}
K(0\oplus 1)
&= \{ u\in U \mid u\succeq 0, 0\oplus 1 \preceq \ker(u) \} \\
&= \{ \eta\cdot [(-2\Re\,\tilde z-1)p(-\tilde z) 
+ (-2\Re\,\tilde z+1)p(\tilde z)] \mid \\
& \hspace{15em} |\tilde z|=1, \Re(\tilde z)\leq-1/2, \eta\geq 0 \} \oplus 0\\
&= \{ \Re(\tilde z)(\sigma_X-2\cdot\id)+\Im(\tilde z)\sigma_Y \mid 
\arg(\tilde z)\in\,[\tfrac{2}{3}\pi,\tfrac{4}{3}\pi]\}\oplus 0
\end{align*}
has dimension two. Since $\arg(z_+)=\tfrac{2}{3}\pi$ and 
$\arg(z_-)=\tfrac{4}{3}\pi$, we have
\[
\Re(z_\pm)(\sigma_X-2\cdot\id)+\Im(z_\pm)\sigma_Y
=\id+\left(\begin{smallmatrix}0 & \bar z_\pm \\ z_\pm & 0\end{smallmatrix}\right)
=2p(z_\pm)
=u_\pm.
\]
So $K(p_\pm)$ are the extreme rays of $K(0\oplus 1)$ in agreement
with Theorem~\ref{thm:ray-coatoms}(2).
%
%%%%%%%%%%%%%%%%%%%%%%%%%%%%%%%%%%%%%%%%%%%%%%%%%%%%%%%%%%%%%%%%%%%%%%%%%%%%
%%%%%%%%%%%%%%%%%%%%%%%%%%%%%%%%%%%%%%%%%%%%%%%%%%%%%%%%%%%%%%%%%%%%%%%%%%%%
%%%%%%%%%%%%%%%%%%%%%%%%%%%%%%%%%%%%%%%%%%%%%%%%%%%%%%%%%%%%%%%%%%%%%%%%%%%%
%%%%%%%%%%%%%%%%%%%%%%%%%%%%%%%%%%%%%%%%%%%%%%%%%%%%%%%%%%%%%%%%%%%%%%%%%%%%
%%%%%%%%%%%%%%%%%%%%%%%%%%%%%%%%%%%%%%%%%%%%%%%%%%%%%%%%%%%%%%%%%%%%%%%%%%%%
%
\section{Many-body systems}
\label{sec:Composite}
\par
In this section we discuss $k$-local Hamiltonians
and quantum marginals.
\par
We consider a composite system of $N\in\bN$ units, labeled by
$[N]:=\{1,\ldots,N\}$. For each unit $i\in[N]$ we choose a Hilbert space 
$\bC^{n_i}$, $n_i\in\bN$, and a *-algebra $\cA_i$ acting on $\bC^{n_i}$
and containing the $n_i$-by-$n_i$ identity matrix. For any subset 
$\nu\subset[N]$, the tensor product $\bigotimes_{i\in\nu}\bC^{n_i}$ is
the Hilbert space and $\cA_\nu:=\bigotimes_{i\in\nu}\cA_i$ is the 
algebra, with identity denoted by $\id_\nu$, of the system composed of 
the units in $\nu$. The full system has algebra 
\[\textstyle
\cA:=\cA_{[N]}=\bigotimes_{i\in[N]}\cA_i.
\]
By definition, a {\em $k$-local Hamiltonian} is of the form
\begin{equation}\textstyle\label{eq:k-local}
a=\sum_{|\nu|=k}a(\nu)\otimes\id_{\nu'},
\end{equation}
where $\nu'=[N]\setminus\nu$, where $a(\nu)\in\cA_\nu$ is a hermitian 
matrix, and where the sum extends over subsets $\nu\subset[N]$ with 
$|\nu|=k$. We denote the real vector space of $k$-local Hamiltonians 
by $U_{(k)}$.
\par
From now on we consider $U=U_{(k)}$. We pointed out in the introduction
that $\pi(C)$ is isomorphic to the set of $k$-body marginals. To see 
this, let $\nu\subset[N]$. The {\em partial trace} over the subsystem 
$\nu$ is the linear map $\tr_{\nu}:\cA\to\cA_{\nu'}$ which is the adjoint 
of the embedding $\cA_{\nu'}\hookrightarrow\cA$, $a\mapsto a\otimes\id_\nu$ 
with respect to the Hilbert-Schmidt inner product. If $\rho\in C_\cA$ is 
a state on $\cA$ then $\tr_{\nu'}(\rho)\in C_{\cA_\nu}$ is the 
{\em marginal} of $\rho$ on subsystem $\nu$. For $k=1,\ldots,N$ we define 
the linear map
\begin{equation}\textstyle\label{eq:par-tr}
\tr_{(k)}:\cA\to\bigtimes_{|\nu|=k}\cA_\nu,
\quad
a\mapsto (\tr_{\nu'}(a))_{|\nu|=k}
\end{equation}
to the cartesian product (direct sum) of algebras $\cA_\nu$, $|\nu|=k$.
The set of $k$-{\em body marginals} or 
$k$-{\em body reduced density matrices} \cite{Chen-etal2012a} is
\begin{equation}\textstyle\label{eq:map-k-body}
\cD_{(k)}:=\{\tr_{(k)}(\rho)\mid \rho\in C_\cA\}.
\end{equation}
The restricted linear map
\begin{equation}\textstyle\label{eq:iso-k-body}
\tr_{(k)}|_{\pi(C_\cA)}:\pi(C_\cA)\to\cD_{(k)}
\end{equation}
is a bijection because of $\tr_{(k)}=\tr_{(k)}\circ\pi$ and since 
$\tr_{(k)}|_{U_{(k)}}$ is injective. 
\par
The inclusion $\cD_{(k)}\subset\bigtimes_{|\nu|=k}C_{\cA_\nu}$ is 
strict for $k\geq 2$ as it follows from comparison of dimensions. Let
$n_1=\cdots=n_N$ and $\cA_i=M_{n_1}$, $i\in[N]$. Then 
Proposition~1 of \cite{Weis-etal2015} and (\ref{eq:iso-k-body}) show
\[\textstyle
\dim(\cD_{(k)})
=\dim(\pi(C_\cA))
=\dim(U_{(k)})-1
=\sum_{\ell=1}^k{N\choose\ell}(n_1^2-1)^\ell.
\]
For $k=2$ and three qubits this gives
$\dim(\cD_{(k)})=36<45=\dim(C_{M_2}\times C_{M_2}\times C_{M_2})$.
\par
In the classical case (\ref{eg:classical}), the algebra of unit $i\in[N]$ 
is the space of functions $\cA_i=\bC^{X_i}$ on a configuration space $X_i$
with $|X_i|=n_i$, $i\in[N]$. The algebra $\cA_\nu=\bC^{X_{\nu}}$ is the
space of functions on the configuration space of subsystem $\nu$,
\begin{equation}\textstyle\label{eq:config-many}
X_{\nu}:=\bigtimes_{i\in\nu}X_i,
\qquad\nu\subset[N].
\end{equation}
The full configuration space is $X:=X_{[N]}$. Viewing $x\in X$ as a sequence,
let $x_\nu\in X_\nu$ denote its truncation to $\nu\subset[N]$. This means 
$x=(x_i)_{i\in[N]}$ for $x_i\in X_i$, $i\in[N]$, and 
$x_\nu=(x_{\nu,i})_{i\in\nu}$ for $x_{\nu,i}=x_i$, $i\in\nu$. Denoting the
disjoint union by
\[\textstyle
\bigcup_{|\nu|=k}X_\nu=\{(\nu,y):|\nu|=k,y\in X_\nu\},
\]
the matrix of (\ref{eq:par-tr}) with respect to the bases 
$(\delta_x)_{x\in X}$ and 
$(\delta_{(\nu,y)})_{(\nu,y)\in\bigcup_{|\nu|=k}X_\nu}$ is
\begin{equation}\textstyle\label{eq:matrix}
(\tr_{(k)})_{(\nu,y),x}
=\left\{\begin{array}{ll} 
0 & \mbox{if $y\neq x_\nu$}\\
1 & \mbox{if $y=x_\nu$}
\end{array}\right.,
\qquad
x\in X,|\nu|=k,y\in X_\nu.
\end{equation}
The simplex $C_\cA=\Delta_X$ is the convex hull of 
$(\delta_x)_{x\in X}$. Hence (\ref{eq:map-k-body}) shows that $\cD_{(k)}$ 
is the convex hull of the columns of the matrix (\ref{eq:matrix}). The 
polytope $\cD_{(k)}$ is well-known in mathematical statistics 
\cite{DevelinSullivant2003,Geiger-etal2006,Kahle2010}. If $n_1=\cdots=n_N$, 
then Proposition~1 of \cite{Weis-etal2015} and (\ref{eq:iso-k-body}) show
\begin{equation}\textstyle\label{eq:2loc3bitDIMgen}
\dim(\cD_{(k)})
=\dim(\pi_{U_{(k)}}(\Delta_X))
=\dim(U_{(k)})-1
=\sum_{\ell=1}^k{N\choose\ell}(n_1-1)^\ell.
\end{equation}
%
%%%%%%%%%%%%%%%%%%%%%%%%%%%%%%%%%%%%%%%%%%%%%%%%%%%%%%%%%%%%%%%%%%%%%%%%%%%%
%%%%%%%%%%%%%%%%%%%%%%%%%%%%%%%%%%%%%%%%%%%%%%%%%%%%%%%%%%%%%%%%%%%%%%%%%%%%
%%%%%%%%%%%%%%%%%%%%%%%%%%%%%%%%%%%%%%%%%%%%%%%%%%%%%%%%%%%%%%%%%%%%%%%%%%%%
%%%%%%%%%%%%%%%%%%%%%%%%%%%%%%%%%%%%%%%%%%%%%%%%%%%%%%%%%%%%%%%%%%%%%%%%%%%%
%%%%%%%%%%%%%%%%%%%%%%%%%%%%%%%%%%%%%%%%%%%%%%%%%%%%%%%%%%%%%%%%%%%%%%%%%%%%
%
\section{A simple three-bit example}
\label{sec:example-3bit}
\par
We discuss ground spaces of $2$-local $3$-bit Hamiltonians.
\par
The three-bit configuration space (\ref{eq:config-many}) is 
$X=\{0,1\}\times\{0,1\}\times\{0,1\}$. The function
\[\textstyle
f:X\to\bR,
\quad
f(x_1,x_2,x_3)=(-1)^{x_1+x_2+x_3}
\]
spans the complement of $U_{(2)}$ in $A=\bR^X$, as $\dim(U_{(2)})=7$ 
by (\ref{eq:2loc3bitDIMgen}). Let $p\in\cP_\cA$. Lemma~\ref{lem:linear} 
shows $K(p)=\{u\in L(p)\mid u\succeq 0\}$ where 
\begin{equation}\textstyle\label{eq:L-for-3bit}
L(p)=p'Ap'\cap U_{(2)}=(pAp+\bR\cdot f)^\perp.
\end{equation}
\par
Let $|p|=7$. We prove $p\not\in\cP(U_{(2)})$. Since $f$ has full rank 
and $p'$ only one non-zero coefficient, for all $a\in p'Ap'$ the 
statement $\langle a,f\rangle=0$ is equivalent to $a=0$. This proves 
$K(p)=\{0\}$. Theorem~\ref{thm:char-PU-K} shows $p\not\in\cP(U_{(2)})$
because $K(\id)=\{0\}$.
\par
Let $|p|\leq 5$. We prove that $p$ is not a coatom of $\cP(U_{(2)})$.
First, let us assume that for all configurations $x\in p'$ we have 
$f(x)=1$ (the case $f(x)=-1$ is analogous). A diagonal matrix $a\in A$ 
lies in $K(p)$ if and only if $a\in p'Ap'$, $a\succeq 0$, and $a\in U$.
Assuming the latter, we obtain $0=\langle a,f\rangle=\tr(a)$ which 
implies $a=0$. This proves $K(p)=\{0\}$ and shows $p\not\in\cP(U_{(2)})$ 
as in the preceding paragraph. Second, assume without loss of generality
that there are $x,y\in p'$ such that $f(x)=1$ and $f(y)=-1$. Since 
$|p'|\geq 3$, there is $z\in p'\setminus\{x,y\}$ such that $f(z)=-1$
(the case $f(z)=1$ is analogous). Using the basis (\ref{eq:delta}) we 
obtain
\[
\delta_x+\delta_y, \delta_x+\delta_z \in K(p)
\]
which shows $\dim\,K(p)>1$. Theorem~\ref{thm:ray-coatoms}(1) shows that 
$p$ is not a coatom of $\cP(U_{(2)})$.%
\newlength{\coatomh}%
\setlength{\coatomh}{2.3cm}%
\begin{figure}
\includegraphics[height=\coatomh]{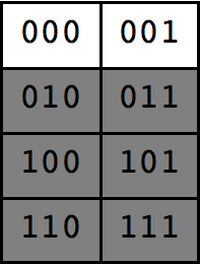}
\includegraphics[height=\coatomh]{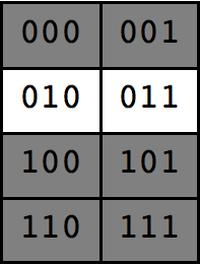}
\includegraphics[height=\coatomh]{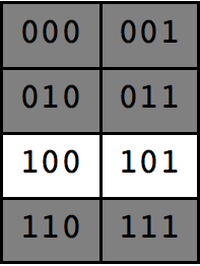}
\includegraphics[height=\coatomh]{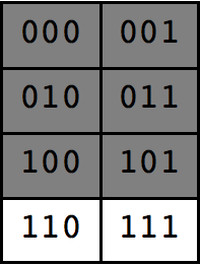}
\includegraphics[height=\coatomh]{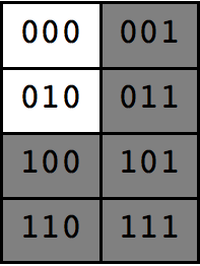}
\includegraphics[height=\coatomh]{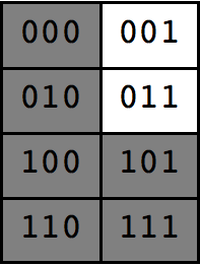}
\includegraphics[height=\coatomh]{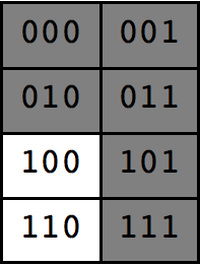}
\includegraphics[height=\coatomh]{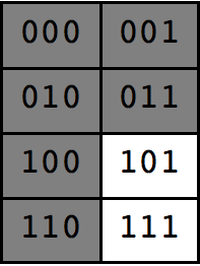}\\[2mm]

\includegraphics[height=\coatomh]{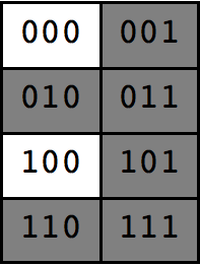}
\includegraphics[height=\coatomh]{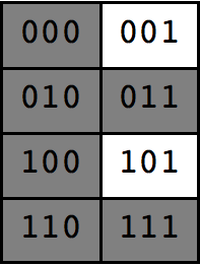}
\includegraphics[height=\coatomh]{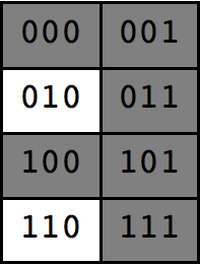}
\includegraphics[height=\coatomh]{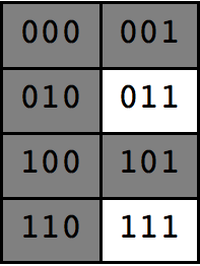}
\includegraphics[height=\coatomh]{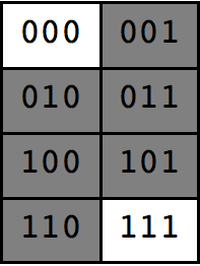}
\includegraphics[height=\coatomh]{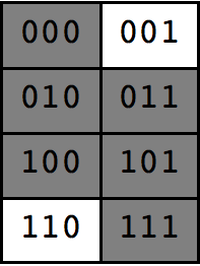}
\includegraphics[height=\coatomh]{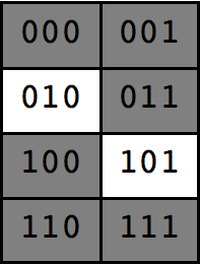}
\includegraphics[height=\coatomh]{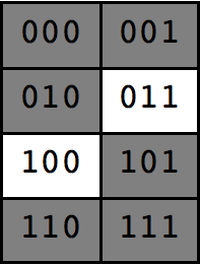}
\caption{\label{fig:coatoms} 
The tables depict the $16$ coatoms of $\cP(U_{(2)})$. Ground 
configurations are distinguished by a dark background.}
\end{figure}%
\par
Let $|p|=6$ and let $p'=\{x,y\}$ for distinct $x,y\in X$. 
If $f(x)=f(y)$ then $K(p)=\{0\}$ follows as in the case $|p|\leq 5$ 
treated above. If $f(x)\neq f(y)$ then $a\in p'Ap'$ and $a\in U$
imply that $a$ is a scalar multiple of $\delta_x+\delta_y$, so
\[
K(p)
=\{u\in U \mid u\succeq 0, p\preceq\ker(u)\}
=\{\eta\cdot(\delta_x+\delta_y) \mid \eta\geq 0\}.
\]
This shows that $p$ is a coatom of $\cP(U_{(2)})$. See
Figure~\ref{fig:coatoms} for drawings.
\par
The complete bipartite graph $K_{4,4}$ with vertex set $X$ and 
bi-partition 
\[
V_i=\{ x\in X \mid f(x)=i\}, \qquad i=\pm1,
\] 
allows to have further insights into the lattice $\cP(U_{(2)})$,
for three bits. We showed that $p\subset X$ is a coatom of 
$\cP(U_{(2)})$ if and only if $p'$ is an edge of $K_{4,4}$. Dually, 
$p\subset X$ is an atom of the dual lattice
\[\textstyle
\cP(U_{(2)})^*=\{p'\mid p\in \cP(U_{(2)}) \}
\]
if and only if $p$ is an edge of $K_{4,4}$. Since $\cP(U_{(2)})$ is 
coatomistic with infimum the intersection (Lemmas~\ref{lem:coatomistic} 
and~\ref{lem:PU}), the dual lattice $\cP(U_{(2)})^*$ is atomistic with 
supremum the union. Thus $p\subset X$ lies in $\cP(U_{(2)})^*$ if and
only if $p$ is a union (possibly empty) of edges of $K_{4,4}$.
\par
It is not clear that any $p\subset X$ with $|p|\geq 5$ lies in 
$\cP(U_{(2)})^*$ because $p\cap V_{+1}\neq\emptyset$ and 
$p\cap V_{-1}\neq\emptyset$. Similarly, $68$ of the $70$ subsets of $X$ 
of cardinality four lie in $\cP(U_{(2)})^*$, the exceptions are $V_{+1}$ 
and $V_{-1}$. Dually, $\cP(U_{(2)})$ contains all $p\subset X$ with 
$|p|\leq 3$, which is a special case of Theorem~14 of \cite{Kahle2010}. 
\par
We discuss ground spaces of frustration-free Hamiltonians
\cite{Beaudrap-etal2010,Movassagh-etal2010,Ji-etal2011,Chen-etal2012a,
DarmawanBartlett2014}, omitting detailed proofs. A $k$-local Hamiltonian 
$a\in U_{(k)}$ is {\em frustration-free} if $a$ has a sum representation 
(\ref{eq:k-local}) such that every ground state of $a$ is a ground state 
of $a(\nu)\otimes\id_{\nu'}$, $|\nu|=k$. The set $U_{(k)}^{\rm ff}$ of 
frustration-free $k$-local Hamiltonians is not a vector space. But the 
set of its ground projections, combined with the zero matrix, is a complete 
lattice, denoted $\cQ(U_{(k)})$, whose infimum is the intersection of 
images (\ref{eq:iso-im}). The lattice is coatomistic, as the ground space 
of $a\in U_{(k)}^{\rm ff}$ is the intersection of ground spaces of the 
local terms $a(\nu)\otimes\id_{\nu'}$, $|\nu|=k$.
\par
The dual lattice $\cQ(U_{(k)})^*$ is atomistic. In the commutative case 
(\ref{eg:classical}) the supremum of $\cQ(U_{(k)})^*$ is the union. For 
three bits, it is easy to see that the atoms of $\cQ(U_{(2)})^*$ are the 
non-horizontal edges of $K_{4,4}$ in the vertex arrangement
\begin{center}
\begin{tabular}{ccc}
$V_{+1}$ & \hspace{2em} & $V_{-1}$ \\\hline
$(000)$ && $(111)$ \\
$(011)$ && $(100)$ \\
$(101)$ && $(010)$ \\
$(110)$ && $(001)$
\end{tabular}.
\end{center}
The lattice $\cQ(U_{(2)})^*$ contains any $p\subset X$ with $|p|\geq 6$, 
because $p\cap V_{+1}$ and $p\cap V_{-1}$ contain at least two points each. 
The lattice $\cQ(U_{(2)})^*$ contains $48$ of the $56$ subsets $p\subset X$ 
with $|p|=5$, the eight subsets containing $V_{+1}$ or $V_{-1}$ are missing. 
Dually, $\cQ(U_{(2)})$ contains all $p\subset X$ with $|p|\leq 2$. The last
observation follows also from Lemma~2 of \cite{Weis-etal2015}, since 
there is a one-to-one correspondence between ground spaces of 
$U_{(k)}^{\rm ff}$ and support sets of probability distributions which, 
in the sense of \cite{Geiger-etal2006}, factor according to the 
log-linear model with generators $\{\nu\subset[N]:|\nu|=k\}$.
%
%%%%%%%%%%%%%%%%%%%%%%%%%%%%%%%%%%%%%%%%%%%%%%%%%%%%%%%%%%%%%%%%%%%%%%%%%%%%
%%%%%%%%%%%%%%%%%%%%%%%%%%%%%%%%%%%%%%%%%%%%%%%%%%%%%%%%%%%%%%%%%%%%%%%%%%%%
%%%%%%%%%%%%%%%%%%%%%%%%%%%%%%%%%%%%%%%%%%%%%%%%%%%%%%%%%%%%%%%%%%%%%%%%%%%%
%%%%%%%%%%%%%%%%%%%%%%%%%%%%%%%%%%%%%%%%%%%%%%%%%%%%%%%%%%%%%%%%%%%%%%%%%%%%
%%%%%%%%%%%%%%%%%%%%%%%%%%%%%%%%%%%%%%%%%%%%%%%%%%%%%%%%%%%%%%%%%%%%%%%%%%%%
%
\section{Conclusion}
\label{sec:conclusion}
\par
We proved two theorems concerning the lattice $\cP(U)$ of ground projections 
of a vector space $U$ of hermitian matrices. First, Theorem~\ref{thm:char-PU-K}
offers an equivalent form of the decision problem of whether a projection 
$p\in\cA$ lies in $\cP(U)$, see Section~\ref{sec:intro}. Second, 
Theorem~\ref{thm:ray-coatoms} proves useful, in Section~\ref{sec:example-3bit}, 
to identify coatoms of $\cP(U)$ for $3$-bit Hamiltonians. Are the theorems 
useful in the non-commutative domain?
\par
We are unable to decisively answer the question, but two remarks are
in place. First, there does not exist any matrix
in the space $U_{(2)}$ of $2$-local $3$-qubit Hamiltonians whose ground 
space has dimension seven. This is provable by showing the infeasibility 
of the real polynomial system from Remark~\ref{rem:top-to-bottom-minors} 
for $k=7$ (algebraic degree two). The \texttt{FindInstance} command of 
Wolfram Mathematica solves the problem in less than half a day on a 1.3 
GHz Intel Core i5 processor, using the cylindrical decomposition 
algorithm \cite{Basu-etal2006}. Second, an interesting problem to study is
whether for every projection $p\in M_8$ of rank ${\rm rk}(p)\leq 5$ either 
$K(p)=\{0\}$ or $\dim\,K(p)>1$ holds. If true, this would imply, with the 
help of the Theorems~\ref{thm:char-PU-K} and~\ref{thm:ray-coatoms}, that 
all coatoms of $\cP(U_{(2)})$ have rank six, just as in the commutative 
case of three bits.
%
%%%%%%%%%%%%%%%%%%%%%%%%%%%%%%%%%%%%%%%%%%%%%%%%%%%%%%%%%%%%%%%%%%%%%%%%%%%%
%%%%%%%%%%%%%%%%%%%%%%%%%%%%%%%%%%%%%%%%%%%%%%%%%%%%%%%%%%%%%%%%%%%%%%%%%%%%
%%%%%%%%%%%%%%%%%%%%%%%%%%%%%%%%%%%%%%%%%%%%%%%%%%%%%%%%%%%%%%%%%%%%%%%%%%%%
%%%%%%%%%%%%%%%%%%%%%%%%%%%%%%%%%%%%%%%%%%%%%%%%%%%%%%%%%%%%%%%%%%%%%%%%%%%%
%%%%%%%%%%%%%%%%%%%%%%%%%%%%%%%%%%%%%%%%%%%%%%%%%%%%%%%%%%%%%%%%%%%%%%%%%%%%
%
\vspace{1\baselineskip}
{\par\noindent\footnotesize
{\em Acknowledgements.}
This article was put together 2016--2018 while the author was a postdoctoral
scholar at IMECC, Unicamp, Brazil, 
and at QuIC, Université libre de Bruxelles, Belgium. 
He is glad of having been selected for a grant to join the 
trimester Analysis in Quantum Information Theory at 
Institut Henri Poincaré, Paris, France, in 2017. 
Among others, he thanks  
Alihu{\'e}n Garc{\'\i}a Pavioni, 
Arleta Szko{\l}a, 
Aurelian Gheondea,
Chi-Kwong Li,
Eduardo Garibaldi,
Frederic Shultz, 
Ilya Spitkovsky, 
Ivan Todorov,
Jérémie Roland,
Marcelo Terra Cunha,
Michael Walter, 
Micha{\l} Horodecki,
Pawe{\l} Horodecki, 
Ramis Movassagh, and
Toby Cubitt
for discussions concerning the article, 
lattices of projections, 
the quantum marginal problem, 
and ground spaces.}
%
%
%%%%%%%%%%%%%%%%%%%%%%%%%%%%%%%%%%%%%%%%%%%%%%%%%%%%%%%%%%%%%%%%%%%%%%%%%%%%
%%%%%%%%%%%%%%%%%%%%%%%%%%%%%%%%%%%%%%%%%%%%%%%%%%%%%%%%%%%%%%%%%%%%%%%%%%%%
%%%%%%%%%%%%%%%%%%%%%%%%%%%%%%%%%%%%%%%%%%%%%%%%%%%%%%%%%%%%%%%%%%%%%%%%%%%%
%%%%%%%%%%%%%%%%%%%%%%%%%%%%%%%%%%%%%%%%%%%%%%%%%%%%%%%%%%%%%%%%%%%%%%%%%%%%
%%%%%%%%%%%%%%%%%%%%%%%%%%%%%%%%%%%%%%%%%%%%%%%%%%%%%%%%%%%%%%%%%%%%%%%%%%%%
%
\vspace{1\baselineskip}
\par\noindent
Stephan Weis\\
e-mail: \texttt{maths@weis-stephan.de}\\[.5\baselineskip]
Centre for Quantum Information and Communication\\
Ecole Polytechnique de Bruxelles\\
Universit\'e libre de Bruxelles \\
50 av.~F.D. Roosevelt \\
1050 Bruxelles\\
Belgium
%
%%%%%%%%%%%%%%%%%%%%%%%%%%%%%%%%%%%%%%%%%%%%%%%%%%%%%%%%%%%%%%%%%%%%%%%%%%%%
%%%%%%%%%%%%%%%%%%%%%%%%%%%%%%%%%%%%%%%%%%%%%%%%%%%%%%%%%%%%%%%%%%%%%%%%%%%%
%%%%%%%%%%%%%%%%%%%%%%%%%%%%%%%%%%%%%%%%%%%%%%%%%%%%%%%%%%%%%%%%%%%%%%%%%%%%
%%%%%%%%%%%%%%%%%%%%%%%%%%%%%%%%%%%%%%%%%%%%%%%%%%%%%%%%%%%%%%%%%%%%%%%%%%%%
%%%%%%%%%%%%%%%%%%%%%%%%%%%%%%%%%%%%%%%%%%%%%%%%%%%%%%%%%%%%%%%%%%%%%%%%%%%%
%
\bibliographystyle{plain}

\end{document}